\documentclass[12pt]{article}
\usepackage{amssymb,amsmath,cite}
\long\def\@makefntext#1{\parindent 0cm\noindent
\hbox to 1em{\hss$^{\@thefnmark}$}#1}

\begin{document}
\begin{titlepage}
\vspace{.5in}
\begin{flushright}
March 2007\\
\end{flushright}
\vspace{.5in}
\begin{center}
{\Large\bf
 Symmetries, Horizons,\\[.8ex] and Black Hole Entropy}\\
\vspace{.4in}
{S.~C{\sc arlip}\footnote{\it email: carlip@physics.ucdavis.edu}\\
       {\small\it Department of Physics}\\
       {\small\it University of California}\\
       {\small\it Davis, CA 95616}\\{\small\it USA}}
\end{center}

\vspace{.5in}
\begin{center}
{\large\bf Abstract}
\end{center}
\begin{center}
\begin{minipage}{4.7in}
{\small
Black holes behave as thermodynamic systems, and a central task 
of any quantum theory of gravity is to explain these thermal 
properties.  A statistical mechanical description of black hole 
entropy once seemed remote, but today we suffer an embarrassment 
of riches: despite counting very different states, many 
inequivalent approaches to quantum gravity obtain identical results.  
Such ``universality'' may reflect an underlying two-dimensional 
conformal symmetry near the horizon, which can be powerful enough 
to control the thermal characteristics independent of other details 
of the theory.  This picture suggests an elegant description of the 
relevant degrees of freedom as Goldstone-boson-like excitations 
arising from symmetry breaking by the conformal anomaly.
}
\end{minipage}
\end{center}
\end{titlepage}
\addtocounter{footnote}{-1}

\section*{The Problem of Universality}

Black holes are thermal systems, radiating as black bodies with characteristic 
temperatures and entropies. Classically, this behavior is a mystery: by Wheeler's 
famous dictum, ``black holes have no hair,'' no classical degrees of freedom to 
account for such thermodynamic properties.  The likely explanation is that the 
relevant microscopic degrees of freedom are fundamentally quantum mechanical.  
Indeed, the Bekenstein-Hawking entropy
\begin{equation}
S = \frac{A_{\scriptstyle\mathit horizon}}{4\hbar G}
\label{a1}
\end{equation}
depends upon both Planck's constant $\hbar$ and Newton's constant $G$, 
hinting that black hole thermodynamics unites quantum mechanics and gravity.

Until recently, little was known about such quantum degrees of freedom.  
Today, we suffer an embarrassment of riches.  Black hole thermodynamics 
can be explained by
\begin{itemize}\renewcommand{\labelitemi}{\labelitemii}
\item weakly coupled string and D-brane states \cite{StromVafa};
\item states of a dual conformal field theory ``at infinity'' \cite{Skenderis};
\item horizonless ``fuzzball'' geometries \cite{Mathur};
\item spin network states at \cite{Ashtekar} or inside \cite{Livine} the 
  horizon;
\item ``heavy'' degrees of freedom in induced gravity \cite{Fursaev};
\item elements of a causal set in the horizon's domain of dependence 
  \cite{Rideout};
\item inherently global characteristics \cite{Hawkingb};
\item entanglement entropy of quantum fields across the horizon 
 \cite{Bombelli,Emparanb}.
\end{itemize}
None of these accounts is complete; string theory, for instance, is most
reliable for supersymmetric black holes, while the loop quantum gravity 
calculations depend on the poorly understood Immirzi parameter.  But 
within their domains of applicability, all seem to work.  

This ``problem of universality'' is already present within particular
models.  The simplest string theory approach, for example, counts 
weakly coupled brane configurations.  But this computation does not yield 
the Bekenstein-Hawking formula (\ref{a1}) directly; rather, one must 
separately determine the entropy and the horizon area in terms of a set 
of charges, and then check, case by case, that they match.  Similarly, 
loop quantum gravity uses special features of four-dimensional spacetimes; 
it works for many different (3+1)-dimensional black holes, but does not 
explain what happens in other dimensions.  More generally, in the 
absence of classical degrees of freedom to which one could apply the 
correspondence principle, it is not clear why \emph{any} counting of 
microstates should reproduce Hawking's original results \cite{Hawking}, 
which were based on quantum field theory in a fixed, classical gravitational 
background.

\section*{Symmetries and Microstates}

We do not know why such disparate computations yield the same black hole 
entropy.  But one intriguing possibility is that a classical symmetry 
near the horizon may govern the number of degrees of freedom, independent of 
the details of quantum gravity.  

To an observer outside a black hole, the 
near-horizon region is effectively two-dimensional and conformally invariant 
\cite{Birmingham,Medved}: transverse excitations and dimensionful quantities 
are red-shifted away relative to the degrees of freedom in the $r$-$t$ plane.
Thermodynamic quantities such as temperature are conformally invariant 
\cite{Jacobson}; indeed, a conformal description is powerful enough to 
determine the flux \cite{Wilczek} and spectrum \cite{Iso} of Hawking 
radiation.

Remarkably, such a symmetry may also provide a universal explanation 
of black hole entropy.  As Cardy has shown \cite{Cardy}, the entropy of a 
two-dimensional conformal field theory is completely fixed by a few
parameters determined by the symmetry.  More precisely, the fundamental 
symmetry of a two-dimensional conformal field theory is invariance under 
holomorphic coordinate transformations $z\rightarrow z + \xi(z)$, 
${\bar z}\rightarrow {\bar z} + {\bar\xi}({\bar z})$.  The Poisson brackets 
of the generators of such transformations form a Virasoro algebra
\begin{align}
\left\{L[\xi],L[\eta]\right\} &= L[\xi{\dot\eta}-\eta{\dot\xi}]
  + \frac{c}{48\pi}\int dt ({\dot\xi}{\ddot\eta}-{\dot\eta}{\ddot\xi}) ,
\label{a2}
\end{align}
with a corresponding expression for ${\bar L}[{\bar\xi}]$.  The first
term on the right-hand side gives the ordinary commutator of vector fields.  
The second is the unique central extension, completely characterized by the 
``central charge'' $c$.  Like other symmetries, conformal invariance implies 
the existence of conserved charges: the zero-mode generators $L_0 = L[\xi_0]$ 
and ${\bar L}_0 = {\bar L}[{\bar\xi}_0]$ are ``conformal charges'' roughly 
analogous to energies. 

Now consider any two-dimensional conformal field theory for which the lowest 
eigenvalues $\Delta_0$ and ${\bar\Delta}_0$ of $L_0$ and ${\bar L}_0$ 
are nonnegative.  Cardy's striking result is that the asymptotic density 
of states at large conformal charge $(\Delta,\bar\Delta)$ takes the simple 
form
\begin{equation}
\ln\rho(\Delta,{\bar\Delta}) \sim 
  2\pi\sqrt{\frac{(c-24\Delta_0)\Delta}{6}} 
  + 2\pi\sqrt{\frac{({\bar c}-24{\bar\Delta}_0){\bar\Delta}}{6}} \ ,
\label{a3}
\end{equation}
independent of any other details of the theory.  This is precisely the kind of
universal behavior we need.

\section*{How to Ask the Right Question}

Before proceeding further, we must confront a fundamental problem: how do we
formulate our questions to ensure that we are asking about a black hole?  In 
semiclassical computations, this is not an issue---we simply choose a fixed black 
hole geometry and analyze fields and metric fluctuations in that background.  
In a full quantum theory of gravity, though, we cannot do this---the components 
of the metric do not commute, and cannot be specified simultaneously.  We must 
instead find new conditions strong enough to ensure the presence of a black hole, 
but weak enough to be allowed by quantum mechanics.

The simplest conditions of this sort are restrictions on the asymptotic behavior 
of the metric.  The basic symmetry of general relativity is diffeomorphism 
invariance, manifested in the Hamiltonian formalism through a set of constraints% 
---the ``diffeomorphism constraints'' ${\cal H}_i$ and the ``Hamiltonian constraint'' 
${\cal H}_\perp$---that generate coordinate transformations.  When boundary 
conditions are imposed, these constraints acquire boundary terms, which can 
change their Poisson algebra.  For the (2+1)-dimensional black hole, these
terms lead to Virasoro algebra at infinity \cite{Brown} that gives, via the 
Cardy formula, just the right enumeration of states to explain the 
Bekenstein-Hawking entropy \cite{Carlip}.  In general, though, asymptotic 
conditions are too weak---they cannot distinguish between a black hole and a 
``star''---and the results depend on particular features of (2+1)-dimensional 
spacetime that are not easily generalized.  Many near-extremal black holes 
studied in string theory have near-horizon geometries that look 2+1 dimensional, 
allowing one to apply this method \cite{Skenderis}, but clearly a more general 
approach would be desirable.

One such generalization is to treat the \emph{horizon} as a boundary---or,
more precisely, as a hypersurface upon which we impose ``boundary conditions.''  
Once again, such restrictions alter the symmetry algebra of general relativity.  
Now, in \emph{any} spacetime dimension greater than two, the result is a 
Virasoro algebra with the right central charge and conformal charges to 
yield the Bekenstein-Hawking entropy \cite{Carlipa,Carlipb,Cvitan}.  But the 
diffeomorphisms whose algebra leads to this result are generated by vector 
fields that blow up at the horizon \cite{Dreyer,Koga}, and this divergence 
is poorly understood.   

The newest approach \cite{Carlipd,Carlipe} is to impose the existence
of a horizon as a constraint on initial values of gravitational degrees 
of freedom.  We begin again with standard general relativity, but now add 
a set of ``horizon constraints'' $K_\alpha=0$ on an initial hypersurface.
These may ensure, for example, that a chosen hypersurface has vanishing 
expansion and a prescribed surface gravity.  Such horizon constraints will 
typically fail to commute with the diffeomorphism and Hamiltonian constraints,
but we can cure this in a manner first suggested by Bergmann and Komar 
\cite{Bergmann}: we add ``zero,'' in the form of multiples of the $K_\alpha$, 
to ${\cal H}_i$ and ${\cal H}_\perp$ to produce new generators that commute 
with the $K_\alpha$.  This will change the Poisson algebra of ${\cal H}_i$ 
and ${\cal H}_\perp$, potentially giving rise to the desired central charge.  
In \cite{Carlipe}, this program was carried out for a very general version 
of two-dimensional dilaton gravity, using the ``radial quantization'' techniques
popular in string theory.  The result was again a Virasoro algebra that yielded
the correct Bekenstein-Hawking entropy.  For now, this ``horizon constraint'' 
approach seems the most general; it appears to incorporate both the ``horizon 
as boundary'' results and the method of asymptotic symmetries.

\section*{What Are We Counting?}

The advantage of the Cardy formula is that while it lets us counts states, 
it does not require detailed knowledge of the states being counted.  Nevertheless, 
these results suggest an interesting effective description of black hole entropy.  

In conventional quantum gravity, physical states are required to be invariant 
under diffeomorphisms; in our context,
\begin{equation}
L[\xi]|\mathit{phys}\rangle = {\bar L}[{\bar\xi}]|\mathit{phys}\rangle = 0.
\label{f1}
\end{equation}
If the Virasoro algebra (\ref{a2}) has a central charge, though, such conditions
are incompatible with the Poisson brackets.  We know how to weaken (\ref{f1}):
we can require, for example, that only positive frequency modes annihilate physical 
states \cite{CFT}.  But then new states that had formerly been excluded---for 
instance, the ``descendant states'' $L_{-n}|0\rangle$---become physical.

This phenomenon is strongly reminiscent of the Goldstone mechanism \cite{Kaloper}.  
The conformal anomaly breaks diffeomorphism invariance, and as a consequence,
``would-be pure gauge'' degrees of freedom become physical.  For asymptotically 
anti-de Sitter spacetimes in three \cite{Carlipf} and five \cite{Aros} dimensions, 
an explicit description of the resulting degrees of freedom at infinity is possible; 
one might hope for something similar at a horizon.

Perhaps the most important open question is whether Hawking radiation and black 
hole evaporation can also fit into this approach.  As noted above, 
one can compute Hawking radiation with techniques that rely on conformal anomalies 
of the radiating matter fields \cite{Wilczek,Iso}, and Emparan and Sachs have 
shown that in 2+1 dimensions, a scalar field can be coupled to the conformal 
boundary degrees of freedom to obtain Hawking radiation \cite{Emparan}.  If a 
similar mechanism could be found at the horizon, it would represent major progress.

\vspace{1.5ex}
\begin{flushleft}
\large\bf Acknowledgments
\end{flushleft}

This work was supported in part by Department of Energy grant
DE-FG02-91ER40674.

\end{document}